\documentclass[10pt]{iopart}
\usepackage{amsfonts}
\usepackage{amssymb}

 \eqnobysec
 \begin{document}

 \title{Backwards on Minkowski's road. From $4D$ to $3D$ Maxwellian electromagnetism}

\author{Yakov Itin$^{(1,2)}$ and Y. Friedman$^{(1)}$}

\address{$^{(1)}$Jerusalem College of
    Technology, \\$^{(2)}$Institute of Mathematics, Hebrew University of
   Jerusalem }
\ead{itin@math.huji.ac.il}
    \begin{abstract}
Minkowski's concept of a four-dimensional physical space is a
central paradigm of modern physics. The three-dimensional Maxwellian 
electrodynamics is uniquely generalized to the covariant four-dimensional form. 
Is the (1+3) decomposition of the covariant four-dimensional form
unique? How do the different sign assumptions of electrodynamics emerge
from this decomposition? Which of these assumptions are fundamental
and which of them may be modified?  How does the
Minkowski space-time metric emerge from this preliminary metric-free
construction? In this paper we are looking for answers to the 
problems mentioned. Our main result is the derivation of four different
possible sets of electrodynamic equations which may occur in
different types of isotropic electromagnetic media. The wave propagation 
in each of these media is described by the Minkowskian optical
metrics. Moreover, the electric and magnetic energies are nonnegative in all cases. 
We also show that the correct directions of the Lorentz force (as a consequence of the  
Dufay  and the Lenz rules) hold true for all these cases. 
However, the differences between these four types of media must have a physical meaning. 
In particular, the signs of the three electromagnetic invariants are different.  
    \end{abstract}



    \renewcommand{\leftmark}
     {Y. Itin and Y. Friedman: Backwards on Minkowski's road}





\date{\today}


    \section{Introduction}
    One of the most famous result of Hermann Minkowski in physics
     is the four-dimensional Minkowski metric of space-time \cite{Minkowski}.
 This concept generated
      the  generally adopted form of special relativity and formed a
      basis of general relativity and quantum field theory.
      Apparently in mathematical literature this metric is referred
      to as the Lorentzian metric, i.e, after a  physicist
      Hendrik Lorentz and not after a  mathematician
      Hermann Minkowski.  This strange situation  is due to
      the fact that another principally different and not 
      less famous Minkowski metric is a central concept in the theory
      of final normalized functional spaces.

    Although the Minkowski metric is a fundamental concept of  classical and quantum physics, its origin does not come
    from some general philosophical paradigm. The $4D$
    Minkowski metric has its roots in  Maxwell's 
theory of the electromagnetic field. This fact is already clear from
the original title  of  Einstein's first special relativity
paper: $``$Zur Elektrodynamik bewegter K{$\ddot {\rm o}$}rper" ($``$On the
Electrodynamics of Moving Bodies").

 Consequently, the special choice of signs in the Minkowski metric,
 or, more generally, the signature of the curved spacetime metric,
 has to be in a correlation with various sign conventions and rules
 used in classical three-dimensional Maxwellian electrodynamics.
Recall some of these  assumptions on which the signs are based:
   \begin{itemize}
   \item[i] The special sets of signs in every one of the four Maxwell
   equations when they are written in $3D$-space;
\item[ii] positivity of the electromagnetic energy;
\item[iii] two possibilities of the signs of the electric charges;
\item[iv] attraction between opposite charges and repulsion between
charges of the same sign --- Dufay's rule;
\item[v] pulling of a ferromagnetic core into a solenoid independently
 of the direction of the current --- Lenz's rule;
\item[vi] positivity of the  electric permittivity  and the  magnetic
permeability constants for most of the natural dielectric materials.
 \end{itemize}
The aim of the current paper is to study which of these sign
assumptions are only customarily accepted conventions and which of
them are consequences of the fundamental physical laws and cannot be
modified without changing the laws.  On the other hand, which of
them can be modified without braking the fundamental 4D laws? Do
such modified laws possibly describe some physical meaningful reality, 
electromagnetic behavior in manufactured materials, for instance?

In line of Minkowski's idea  that  3D physics has to be
considered as a chapter of  4D physics,  3D Maxwellian 
electromagnetism  becomes  part of the covariant 4D electromagnetic theory.
Differential forms and tensors are commonly used for such a 
generalization. In this paper, we follow the differential form
approach.

In the standard form of   Maxwell's theory, all the sign
assumptions listed above appear together and the Minkowski metric
is postulated  from the very beginning. Such a construction does not
allow to investigate the specific meaning of the electromagnetic
sign conventions and their relations to the Minkowski signature. In
this paper, we base our consideration on the premetric formulation
of the electrodynamics. Although the roots of such an approach can be
found in  the older literature \cite{Post62}, its final form was derived only recently
in a series of papers \cite{ROH2002}-\cite{HIO} and in a book \cite{book} of Hehl and
Obukhov.

    The current paper is a continuation of the paper \cite{Itin:2004qr}
     due to Hehl and the first-named author.
Instead of using a general constitutive relation constrained only
by the reciprocity relation, as it was made in \cite{Itin:2004qr}, we
restrict ourselves now to the electrodynamics in vacuum and in isotropic media.
For such media we may assume that the metric in the "rest space
hypersurfaces" is Euclidean.  With this restriction we hope to go
deeply into the  study of the roots of the the sign convections listed above.
Our primary aim is to find the physically motivated conditions which
enable the existence of non-ordinary materials with  non-positive
electric permittivity and/or magnetic permeability constants.
 We start with a four-dimensional electromagnetic theory based 
on the following assumptions: 
 \begin{itemize}
   \item[i] Charge conservation;
   \item[ii] flux conservation; 
\item[ii] positivity of the electric and magnetic  energy;
\item[iii] energy-momentum conservation;
\item[iv] local linear constitutive law which we restrict to the  isotropic form.
 \end{itemize}
 
 We show that under these assumptions a unique $4D$ electromagnetic system is reduced 
to four different types of  $3D$ Maxwell-type systems. For each of these models,
the electric and magnetic energies are positive and  the Lorentz force has a standard 
direction (Dufay  and Lenz rules). Moreover the optical metric is Minkowskian for all 
these  models. However, they differ in the values of the four electromagnetic invariants. 
We also briefly discuss the possibility  to identify our models with recently 
manufactured metamaterials. 

    The organization of the paper is as follows:
    In  section 2, we present some mathematical preliminaries
    and notations which are used in the sequel.
In section 3, we give a brief account of  premetric
electrodynamics. Here, our main aim is to introduce an $(1+3)$
decomposition relative to an arbitrary observer.
 These sign factors in Maxwell equations,  which have no physical
 meaning and are only  subject of conventions,  are determined here.
Section 4 is devoted to the constitutive relation. We define a
restricted constitutive map based on transformation properties
of the fields and their different physical dimensions. In the
$(1+3)$ decomposition, we arrive at  isotropic media with the two
electromagnetic constants $\varepsilon$ and $\mu$. In Section 5,
we consider the expression of the energy-momentum current in differential forms. The conservation of the energy-momentum current 
is required to be in conformity with the conservation of the
charge current (the same orientation of the boundary is used).
Consequently, we derive the sign of the Lenz rule. 
Moreover, we come to a list of four different types of  
electromagnetic media. This is our main result. The optical
properties of the media are studied in Section 6 and 
the optical metric of Minkowskian  signature is
derived. In section 7, we consider the Lorentz force expression
and derive the Dufay and Lenz rules for
different media. In Section 8, we give a summary of our results.
Section 9 is devoted to conclusions and a discussion.

\section{Mathematical preliminaries}
    \subsection{Local observer on a premetric manifold}

We start with a bare four-dimensional  differential manifold $M$
 as a preliminary model of   the physical space-time. As mentioned
by Minkowski \cite{Minkowski} $``$We should have in the world no
longer \textit{space}, but an infinite number of spaces,
analogously as in three-dimensional space an infinite number of
planes. Three-dimensional geometry becomes a chapter in
four-dimensional physics". This suggests to introduce a foliation
of the manifold $M$ by a set of  smooth non-intersecting 
hypersurfaces representing the rest space.

Although on $M$ an invariant meaning can be given only to $4D$
tensorial quantities, their real physical nature emerges  when a
notion of a local observer is introduced. As  Minkowski write
 in \cite{Minkowski}: $``$The separation of the field produced by the
 electron into electric and magnetic force is relative with regard
 to the underlying time axis".
 This is similar to the notion of an observer used
in modern literature.

On a manifold endowed with a metric of Minkowski signature, two
different, but locally equivalent, descriptions of an observer are
in use \cite{Bini:1995hx}. Due to the ``congruence point of view", the
observer traveling on a wordline $z^i=z^i(\tau)$ is described
locally by a timelike tangent vector field $n^i=\partial
z^i/\partial\tau$. The orthogonal complement to this local
 field defines the local rest space. In the second
$``$slicing approach", one starts with a foliation of the total
space by spacelike hypersurfaces $\Sigma_\tau$ and reinstates (in
a locally unique manner) the timelike field orthogonal to
$\Sigma_\tau$. It is clear that  in both approaches the metric tensor
plays a crucial role.

Since we are dealing with a manifold without predefined metric structure,
 the notion of a local observer must be modified to involve both of 
 these approaches.

 A {\it local observer} on $M$ is defined by

(i) a foliation of the manifold $M$ generated by a set of  smooth
non-intersecting  hypersurfaces $\Sigma_\tau$, each of which is
diffeomorphic to ${\mathbb R}^3$. The hypersurfaces are
continuously numbered by a real parameter $\tau\in \mathbb R$
and represent the rest space;

(ii) a field of directed curves $z^i=z^i(\tau)$ every one of
which is diffeomorphic to $\mathbb R$ and parametrized by the
same  real parameter $\tau$ with $z^i(\tau)\in \Sigma_\tau$.  The
tangent vector $\mathbf{n}$
  \begin{equation}\label{normal}
    n^i=\frac {\partial z^i}{\partial \tau}
    \end{equation}
is assumed to be defined and differs from zero for every value of
$\tau$;

(iii) every curve $z^i(\tau)$  is is assumed to be transversal to
the hypersurfaces $\Sigma_\tau$ in the following sense: 
Locally, $\Sigma_\tau$ is described by the equation $d\tau =0$,
such that $\mathbf{n}\rfloor d\tau =1$. Here, and in the sequel,
$\rfloor$ denotes the interior product operation.

The parameter $\tau$ will serve as a prototype of a \textit{time}
coordinate while the folio $\Sigma_\tau$ is a prototype of the
\textit{ rest space} for a chosen observer.

In this paper, we will deal with the local properties of the
manifold, so we will restrict to a bounded region $R\subset M$.
Without any restriction on the topological nature of $M$, this
region can be  considered to be orientable. Moreover, we assume
that a defined orientation, i.e., a positive non degenerated
volume element ${}^{(4)}{\rm vol}$ is chosen on $R$.

 Relative to the  chosen observer, the positive volume element
 ${}^{(4)}{ vol}$ is decomposed as
    \begin{equation}\label{vol}
    {}^{(4)}{ vol}=d\tau\wedge {}^{(3)}{ vol}\,.
    \end{equation}
   This equation defines uniquely   a positive volume element
   ${}^{(3)}{vol}$ on a
    hypersurface $\Sigma_\tau$. Hence, a chosen observer necessary
    transforms the positive orientation
     of the total $4D$ space $M$ into a unique positive orientation of
     a $3D$ hypersurface $\Sigma_\tau$.

The introduction of a local observer on $M$ has some implications for
$4D$ differential forms. Since an integration of a differential
form over the oriented submanifold $\Sigma_\tau$ (or over a
domains in it) must yield a non-trivial invariant scalar, the
integrand must be a \textit{twisted} 3-form, i.e.,  such a form 
that changes its sign under basis transformations with negative
determinant. For instance, such a twisted behavior has to be prescribed 
for the electromagnetic current $J$. Since  the electromagnetic
excitation 2-form $H$  is connected to $J$, it  must be a
twisted form, too. 

  Every twisted 3-form lying in a folio has only one component
   which is proportional to the $3D$-volume element. Consequently,
   the notion of a sign can also be rigorously ascribed to such special 3-forms.
   Observe that a general 3-form in four dimensional space has four independent
   components and the positive and negative forms cannot be defined
   rigorously.

On the other hand, the two- and one-dimensional submanifolds of $M$
cannot be given a preferred orientation only by a chosen local
observer. Consequently, an integral over such lower dimensional
submanifolds must involve  \textit{untwisted} differential
forms, i.e., such ones that are not changed under any
transformations of the basis. The electromagnetic field strength
$F$ is an example of such a form.

This simple observation can serve as an additional justification of
the premetric electrodynamics construction which is based on two
integral conservation laws ---  a conservation law for the twisted
3-form $J$ and a conservation law for an untwisted 2-form $F$. In
this sense, the premetric approach is preferable to the ordinary
electrodynamics construction which does not provide us with any
motivation of the different behavior of the two basic
electromagnetic fields.

 \subsection{(1+3)-splitting of differential forms and their differentials}

Consider an arbitrary differential $p$-form $\alpha$ defined on $M$.
Relative to a chosen local observer, the decomposition of this form
is usually given by
\begin{equation}\label{decomp0}
    \alpha=d\tau\wedge \alpha_\bot+\alpha_{||}\,,
    \end{equation}
 where the $(p-1)$-form $\alpha_\bot$ is a  transversal component
 of $\alpha$ while
the $p$-form $\alpha_{||}$  is its longitudinal component.

 Since we are interested not only in a formal decomposition of the
 form $\alpha$, but mainly in a physical interpretation that can
 be given to its parts, we will use a slightly different
 (1+3)-splitting:
    \begin{equation}\label{decomp}
    \alpha=s_{\tt T}d\tau\wedge \beta+s_{\tt S}\gamma\,.
    \end{equation}
 Here we introduced  the sign  factors $s_{\tt T}$ and $s_{\tt S}$
 for the time and  the spatial components, respectively. These sign
 factors have  values from the set $\{-1,+1\}$.   Any
 additional positive scalar factor is assumed to be absorbed in the
corresponding form. The actual rigorous values of the factors
$s_{\tt T},s_{\tt S}$ will to be given in  correspondence with
the physical interpretation of the forms $\beta$ and $\gamma$ and
the physical laws they obey.

The absolute physical dimensions of the $p$-forms $\alpha$ and
$\gamma$ are the same. The $(p-1)$-form $\beta$ has the absolute
dimension of $\alpha$ divided by the dimension of time.
    The forms $\beta$ and $\gamma$ satisfy the relations
     \begin{equation}\label{decomp1}
    \mathbf{n}\rfloor \beta=\mathbf{n}\rfloor \gamma=0\,,
    \end{equation}
    i.e., they  lie in the folio $\Sigma_\tau$.
    Thus the decomposition (\ref{decomp}) is unique and the forms
    $\beta$ and $\gamma$ can be  derived from  $\alpha$ as
    \begin{equation}\label{decomp2}
    \beta=s_{\tt T}\mathbf{n}\rfloor \alpha\,,\qquad \gamma=
    s_{\tt S}\Big(\alpha-s_{\tt T}d\tau\wedge (\mathbf{n}\rfloor \alpha)\Big)\,.
    \end{equation}

     The 4-dimensional exterior derivative operator can be decomposed as
    \begin{equation}\label{dec2}
    d=d\tau\wedge\frac{\partial}{\partial\tau}+\underline{d}\,,
    \end{equation}
    where the spatial exterior derivative $\underline{d}$ refers to
    local coordinates on the
    hypersurface $\Sigma_\tau$. Using the standard formula
for the exterior derivative of a wedge product, we find
    \begin{equation}\label{dec3}
    d\alpha=d\tau\wedge(-s_{\tt T}\,\underline{d}\beta+s_{\tt S}\dot{\gamma})+
    s_{\tt S}\underline{d}\gamma\,.
    \end{equation}
    Here and in the sequel, the partial derivative
    $\partial/\partial\tau$ is abbreviated by a dot on top of the
    corresponding quantity.

    \subsection{Conservation laws in (1+3)-splitting}

For a given $p$-form $\alpha$, a conservation law is described by
the vanishing of the integral
\begin{equation}\label{inegr1}
\int_{\partial C_{p+1}} \alpha=0\,
\end{equation}
over the boundary of any closed  connected
$(p+1)$-dimensional region $ C_{p+1}$. Under this condition,
the Stokes theorem implies
\begin{equation}\label{inegr2}
\int_{\partial C_{p+1}} \alpha=\int_{C_{p+1}} d\alpha=0\,.
\end{equation}
Applying the (1+3)-decomposition (\ref{dec3}) of $d\alpha$,  we
find
\begin{equation}\label{integr6}
\int_{C_{p+1}} d\alpha=s_{\tt S}\int_{C_{p+1}}d\tau\wedge
\left(-\frac{s_{\tt T}}{s_{\tt S}}
\underline{d}\beta+\dot{\gamma}\right)+s_{\tt
S}\int_{C_{p+1}}\underline{d}\gamma=0\,.
  \end{equation}

This equation is  simplified when $\alpha$ is a
$3$-form on a $4$-dimensional manifold $M$. In this case, the
spatial exterior derivative of its longitudinal component is equal
to zero, $\underline{d}\,\gamma=0$. Consequently (\ref{integr6})
takes the form
\begin{equation}\label{integr7}
\frac{s_{\tt T}}{s_{\tt S}}\int_{C_4}d\tau\wedge
\underline{d}\beta=\int_{C_4}d\tau\wedge\dot{\gamma}\,.
\end{equation}
Let  a region be of the form of a tube $C_4=C_3\times[\tau_1,\tau_2]\,$
 with a bounded
hypersurface $C_3$ independent of $\tau$.
When integration over the coordinate $\tau$ is applied on both sides
of (\ref{integr7}), it
can be  rewritten as
\begin{equation}\label{integr7x}
\frac{s_{\tt T}}{s_{\tt S}}\int_{C_3}
\underline{d}\beta=\frac{\partial}{\partial
\tau}\int_{C_3}{\gamma}\,.
\end{equation}
Using once more the Stokes theorem, we can rewrite it as
\begin{equation}\label{integr7xx}
\frac{s_{\tt T}}{s_{\tt S}}\int_{C_2}
\beta=\frac{\partial}{\partial
\tau}\int_{C_3}{\gamma}\,,
\end{equation}
where $C_2=\partial C_3$ is the boundary of $C_3$.

The integrals on both sides of this equation have a clear physical
meaning. On the right hand side we have the time derivative of
the total charge (or of some other physical quantity) denoted by
$\gamma$ contained in the closed region $C_3$. The integral on the left hand side
 represents the flux of the same charge through the
boundary $C_2$ of $C_3$. The sign factor ${s_{\tt T}}/{s_{\tt S}}$
depends now only on the choice of the orientation of the boundary
$C_2$ relative to a given orientation of the region $C_3$. With a
customary choice of the boundary orientation   (the vectors and
1-forms transversal to the boundary are directed outward the
region) we have
\begin{equation}\label{integr8}
\frac{s_{\tt T}}{s_{\tt S}}=-1 \,.
\end{equation}
Under these circumstances,  the (1+3)-decomposition of a
conserved 3-form $\alpha$ has necessarily to be of the form
 \begin{equation}\label{decomp8}
    \alpha=-s_{\tt T}(-d\tau\wedge \beta+\gamma)=
    s_{\tt S}(-d\tau\wedge \beta+\gamma)\,.
    \end{equation}

Moreover, since the 3-form $\gamma$ lies in a folio, it can be
considered as a  positive form (proportional to the volume
element with a positive factor). Hence, the parameter $s_{\tt T}$
supplies a sign to the $3D$-charge $\gamma$ and to the
$4D$-current $\alpha$. Two physically different possibilities are
acceptable:

(i) The charges which can carry only positive sign (like the
energy). In this case we have to choose $s_{\tt S}=1$.

(ii) The charges that can carry both signs (like the electric
charge). In this case the parameter $s_{\tt S}$ can be absorbed
into the form $\gamma$.

In both cases, the decomposition of the form is given by
\begin{equation}\label{decomp9}
    \alpha=-d\tau\wedge \beta+\gamma\,.
    \end{equation}

    If the integral conservation law of a $p$-form $\alpha$ holds
 for an arbitrary closed submanifold $\partial C_{p+1}$, the relation (\ref{inegr2})
  is equivalent to $d\alpha=0$. By using the (1+3)-splitting (\ref{dec3}) of this form, we get
    \begin{equation}\label{dec4}
    d\alpha=0 \qquad \Longleftrightarrow \qquad
    \left\{ \begin{array}{l}
    \underline{d}\beta=-\dot{\gamma}\,,\\ \\
            \underline{d}\gamma =0 \,.\end{array} \right.  
    \end{equation}
   \section{Premetric electrodynamics in (1+3)-splitting}
    \subsection{Electric charge conservation}
    On a $4D$ differential manifold $M$ with a  chosen local observer,
    the electric charge current density $J$ allows to compute the total charge $Q$.
     For this, we have to integrate   over a closed
    oriented $3D$ manifold which is transversal  to the wordline
    of the observer.
     Consequently $J$ must be  given by a  twisted 3-form.
     The decomposition (\ref{decomp}) of a current $J$
    relative to the foliation $\Sigma_\tau$ can be written as
    \begin{equation}\label{decJ}
      J=i_{\tt T}\,d\tau\wedge j+i_{\tt S}\,\rho\,.
    \end{equation}
    Here $\rho$ is a 3-form of the {\it electric charge density} with
    the same absolute dimension as $J$. In the absolute dimensions
    approach \cite{Hehl:2004jn}, we have  $[J]=[\rho]=charge$. 
    The  2-form $j$ is  the {\it electric current density} with
    the absolute dimension of $[j]=charge/time$.
    Observe that both forms lie in the $3D$-folio $\Sigma_\tau$, thus
    the 3-form $\rho$ has only one component,  while  the 2-form
    $j$ has only 3 independent  components.
    Consequently  the decomposition  (\ref{decJ})  is equivalent
    to the ordinary  textbooks description.

    Conservation of  the electromagnetic charge means that the 3-form
    $J$ vanishes, if integrated over an arbitrary closed
    oriented $3D$ submanifold $\partial C_4\in M$, i.e.,
 \begin{equation}\label{charge0}
\oint\limits_{\partial C_4}J=0\,.
\end{equation}
     Since the electric charge can carry two opposite signs, the
    factor $i_{\tt S}$ can be absorbed into the charge form $\rho$ and consequently into 
    $j$. Thus, we assume that $\rho$ can carry two possible signs. Then (\ref{decomp9})
    yields 
    \begin{equation}\label{decJ*x}
      J=-d\tau\wedge j+\rho\,,
    \end{equation}
 which does not involve any undefined sign factors.

    \subsection{Inhomogeneous Maxwell equation}
    In premetric electrodynamics, the inhomogeneous Maxwell equation
    is treated as a consequence of the electric charge conservation law.
    Due to de Rham's theorem, if a 3-form is closed in a region that
    does not admit  non-trivial  3-cycles (cycles which are not boundaries
    of 4D regions), then it is exact.
    Under this condition, the conservation of the electric current
    results in the existence of  a twisted 2-form $H$ of the
    {\it electromagnetic excitation}
    \begin{equation}\label{ax1a}
    dJ=0 \qquad \Longrightarrow \qquad J=dH\,.
    \end{equation}
    Due to this equation, the 2-form  $H$  has the same absolute
    dimension as the current $J$, i.e., $[H]=charge$.

    We write the $(1+3)$-decomposition (\ref{decomp}) of $H$ as
    \begin{equation}\label{decH}
    H=h_{\tt T}\,d\tau\wedge {\mathcal H}+h_{\tt S}\,{\mathcal
    D}\,.
    \end{equation}
    Here ${\mathcal H}$ is the 1-form of the 
    {\it magnetic excitation} with an absolute  dimension $[\mathcal H]
    =charge/time$,
     while ${\mathcal D}$ is the 2-form of the {\it electric  excitation}
     with an absolute dimension $[\mathcal D]=charge$.
     Since both these forms lie in the folio, they have 3 independent
     components, which is consistent with the textbooks description.
We introduced again the sign factors $h_{\tt T}$ and $h_{\tt S}$ with
    values {}from the set $\{+1, -1\}$.

  By applying the exterior derivative operator  to the differential form
  $H$,  we get
 \begin{eqnarray}\label{decHx}
dH&=& d\tau\wedge (-h_{\tt T}\,\underline{d}\,{\mathcal H}+h_{\tt
S}\, \dot{\mathcal D})+h_{\tt S}\,\underline{d}\,\mathcal D\,.
    \end{eqnarray}
Using  (\ref{decJ*x}), we obtain the $1+3$ decomposition of the
equation (\ref{ax1a})
    \begin{equation}\label{inhomMax-0}
dH=J \qquad \Longleftrightarrow \qquad
    \left\{ \begin{array}{l}
 h_{\tt T}\,\underline{d}\,{\mathcal H}-{h_{\tt S}}\,\dot{\mathcal D}=j\,,\\ \\
h_{\tt S}\,\underline{d}\,\mathcal D=\rho \,.\end{array} \right.
    \end{equation}

The second equation , which is  a scalar one in $3D$, can be used to define
the sign of the electric excitation ${\mathcal D}$. Here the
right hand side has a definite sign, which does not change under
coordinate transformation. We assume
 \begin{equation}\label{inhom-sign}
    h_{\tt S}=1\, ,
    \end{equation}
    implying
    \begin{equation}\label{Hdecomp}
      H=h_{\tt T}\,d\tau\wedge {\mathcal H}+\,{\mathcal
    D}\,.
    \end{equation}

Such a choice implies that, for a positive charge, the form
${\mathcal D}$ (similar to a corresponding vector) points out of the
charge.
  Since the first equation is of a vector type,  no
    preferred sign could be associated with $h_{\tt T}$ by a
    choice of a sign of ${\mathcal H}$.  This sign factor remains
    undefined.

    Thus the inhomogeneous Maxwell pair of equations  takes the form
    \begin{equation}\label{inhomMax}
    \left\{ \begin{array}{l}
 h_{\tt T}\,\underline{d}\,{\mathcal H}-\,\dot{\mathcal D}=j\,,\\ \\
\underline{d}\,\mathcal D=\rho \,.\end{array} \right. 
    \end{equation}

    \subsection{Homogeneous Maxwell equation}
    The homogeneous Maxwell equation is dealing with an untwisted 2-form
     $F$ of the {\it electromagnetic field strength}.
    This field is defined by means of    the  Lorentz force density which expresses how the
     electromagnetic field $F$ acts on a test charge with charge density $J$.
    The differential form expression for this force density is \cite{book}
      \begin{equation}\label{axiom2-0}
      {\mathcal F}_\alpha= (e_\alpha\rfloor F) \wedge J\,,
    \end{equation}
    where $e_\alpha$ describes the frame.
    It follows from (\ref{axiom2-0}) that the absolute dimension of
    the field
    $F$ is equal to $[F]=action /charge$,
    which differs from the dimension of the electric excitation $[H]=charge$, see \cite{book}.

   The law of  magnetic flux conservation is given by
    \begin{equation}\label{axiom3}
      \oint\limits_{C_2}F=0\,,
    \end{equation}
    for an arbitrary  closed submanifold $C_2$. Since a two-dimensional
    submanifold of the $4D$-manifold $M$  cannot be
    oriented uniquely by a choice of a local observer,
    the integrand $F$ has to be an untwisted 2-form.
  The equation (\ref{axiom3}) is assumed to hold for an arbitrary submanifold
  $C_2 \subset M$. Thus, it is equivalent to the closure of the form $F$:
 \begin{equation}\label{axiom3x}
    \oint\limits_{C_2}F=0     \qquad \Longleftrightarrow \qquad dF=0\,.
    \end{equation}
    According to (\ref{axiom3x}), the field strength $F$  is determined only up to
 an arbitrary closed 1-form.
However, the expression for the Lorentz force removes this ambiguity.

      The $(1+3)$-decomposition (\ref{decomp}) of the electromagnetic field strength reads
    \begin{equation}\label{axiom2b}
    F=f_{\tt T}\,d\tau\wedge E+f_{\tt S}\,B\,,
    \end{equation}
where $E$ is the  1-form of the {\it electric field strength} while $B$ is the 
2-form of the {\it magnetic field strength.}
 Both forms lie in the folio. Thus, $E$ as well as $B$ has  3 independent components, 
which confirms with the standard description. Also here we
introduced  the sign factors $f_{\tt T}$ and $f_{\tt S}$ with 
values from $\{+1, -1\}$.

By applying  (\ref{dec3}), the
    homogeneous Maxwell equation $dF=0$ decomposes as
    \begin{eqnarray}\label{homMax-1}
      dF&=& d\tau\wedge \left( -f_{\tt
          T}\,\underline{d}\,E+f_{\tt S}\,\dot{B}\right)+f_{\tt
        S}\,\underline{d}\,B=0\,.
    \end{eqnarray}
    Thus,
    \begin{equation}\label{homMax}
     dF=0 \qquad \Longleftrightarrow \qquad
    \left\{ \begin{array}{l}
 f_{\tt T}\,\underline{d}\,E-f_{\tt S}\dot{B} = 0\,,\\ \\
   \underline{d}\,B=0\,. \end{array} \right.  
    \end{equation}
   Due to the homogeneity of these equations, only the quotient  $f_{\tt T}/f_{\tt S}$ can play a
    role. 
Thus one of the factors $f_{\tt T}\,,f_{\tt S}$ is conventional. We choose
\begin{equation}\label{axiom2bx}
    f_{\tt S}=1\, ,
    \end{equation}
    implying
    \begin{equation}\label{Fdecomp}
      F=f_{\tt T}\,d\tau\wedge E+B\,.
\end{equation}
As we will see in the following,  such a choice leads to the usual form of the Lorentz force. 
    In this case, in vacuum or in an ordinary dielectric,
    the force acting on a positive charge is
    directed as the field $E$. We will return to this point below,
    when the (1+3) decomposition of the Lorentz force will be
    considered.

    Left over is  the homogeneous Maxwell pair 
\begin{equation}\label{homMax2}
    \left\{ \begin{array}{l}
 f_{\tt T}\,\underline{d}\,E-\dot{B} = 0\,,\\ \\
   \underline{d}\,B=0\,, \end{array} \right.
    \end{equation}
with one undefined factor $f_{\tt T}$.
    \section{Constitutive relation}
    \subsection{Linear relation and its restrictions}
So far, the electromagnetic fields $F$ and $H$  describe two
different and, at this stage of the 
construction, completely independent physical aspects. The electromagnetic exitation $H$
describes the field generated by the source. The field strength $F$
represent the force acting on a test current. The relation 
between these two fields is a necessary additional element of the
formalism.

Also from the mathematical point of view the system is
undefined. Indeed, the  fields $F$ and $H$  have together 12
components, which are related by 8 independent field equations.

So a {\it constitutive relation} between the fields $F$ and $H$ is
necessary. This relation 
 can be of a rather involved form. For instance,
 in  media with a complicated interior structure (such as a ferromagnet)
  non-linear and non-local constitutive relations are in use.
 Even in the linear case the corresponding tensor has 36
 independent components, which can be decomposed into three irreducible
 pieces.
 In this paper, we will restrict the constitutive relation to its simplest principal 
  part. 
   First we require the operator $\kappa$ to be local and linear.
In this case, a functional constitutive relation takes the form of a
tensorial equation.
   In order to relate an untwisted 2-form to  a twisted one, a constitutive
pseudotensor $\kappa$ has to be involved
\begin{equation}\label{kappa2}
H=\kappa (F)\,.
\end{equation}
 Due to the linearity of the map $\kappa$, the $(1+3)$-decomposition of
(\ref{kappa2}) is given by
\begin{equation}\label{kappa3}
h_{\tt T}\,d\tau\wedge{\mathcal H} +{\mathcal D}
=f_{\tt T}\,\kappa(d\tau\wedge E)+\kappa (B)\,.
\end{equation}
This linear relation  is the most general one. In particular,  the
constitutive pseudotensor $\kappa$ has all its  36 independent
components. In an $(1+3)$-decomposition, these components are arranged
in four $3\times 3$-matrices. Two of these matrices describe
relations between  electric fields and between  magnetic
fields. The two remaining matrices relate an electric field to a magnetic
one and vice versa.  We make now a further principal restriction: 
we assume that the operator $\kappa$ links separately the
electric excitation  $D$ to the electric field strength $E$ and the magnetic
field strength $B$ to the magnetic excitation ${\mathcal H}$. Thus we neglect  the
 magnetoelectric cross effects (like the Faraday effect or
optical activity). In fact, it is known that such effects can
 destroy the light cone structure. Recall, however, that
in this paper we are interested  only in the signature of the
metric, not in its explicit form.

With these restrictions, equation (\ref{kappa3}) splits   into 
\begin{equation}\label{kappa4}
 h_{\tt T}\,d\tau \wedge{\mathcal H} = \kappa (B)\,,\qquad
   {\mathcal D}=f_{\tt T}\,\kappa(d\tau\wedge E)
 \,.
   \end{equation}
   Therefore we have to find an operator $\kappa$ that fulfills  the
following conditions
\begin{equation}\label{const5}
   \kappa(d\tau\wedge E)={f_{\tt T}}{\mathcal D}\,,
    \qquad
    \kappa(B)={h_{\tt T}}d\tau \wedge{\mathcal H}\,.
    \end{equation}
Recall that all the fields involved here  lie in one folio and that both
undefined factors are equal to $\pm 1$.

\subsection{Constitutive pseudotensor and Hodge map}

Roughly speaking, in (\ref{const5}) the operator $\kappa$ has to
perform a sequence of operations:

(i) To remove the timelike element
$d\tau$, i.e., to apply the interior product with the vector
$e_0=\partial/\partial \tau$;

(ii) to transform a 1-form $E$ into a  2-form ${\mathcal D}$.
Since both these form lie in the same folio, it can be done by
applying the three-dimensional Euclidean Hodge map which we will denote by
$\underline{*}$\,;

(iii) we must also take into account that the physical
dimensions of the fields $B$ and ${\mathcal D}$ are different.
Thus we need a dimensional factor, which will be denoted by
$\varphi$.

With these preparations, we can define the action of the operator
$\kappa$ on an arbitrary  2-forms of the type $d\tau\wedge\alpha$
(where $\alpha$ is an arbitrary  1-form lying in the folio) by the relation
   \begin{equation}\label{const7}
   \kappa(d\tau\wedge\alpha )=\varphi \,\underline{*}\alpha\,.
    \end{equation}
We must also  define the action of the operator $\kappa$ on a
 2-forms lying in the folio. Now the operator $\kappa$ has to transform in the folio a
 2-form into a  1-form and to multiply the
result by the time element $d\tau$. The dimensional factor  also
has to be involved. Consequently, for a spacelike 2-form $\beta$,
we define
  \begin{equation}\label{const8}
   \kappa(\beta)=\psi d\tau\wedge \underline{*}\beta\,.
    \end{equation}

This way we have constructed a special constitutive relation for 
an arbitrary 2-form in a $4D$-manifold. Certainly this procedure is
not the only possible one. A much more involved operator can be
introduced \cite{book} even in the local linear case. We assume that
the operator defined by (\ref{const7},\ref{const8}) is a
principal ingredient of every generic constitutive map.

 Since the
3-dimensional Euclidean Hodge map satisfies the relation
$\underline{*}^2=1$, we can derive from (\ref{const7}) and
(\ref{const8})
 \begin{equation}\label{deriv1}
\kappa^2(d\tau\wedge\alpha)=\varphi\kappa(\underline{*}\alpha)=
{\varphi}\psi (d\tau\wedge\underline{*}^2\alpha)= {\varphi}\psi
(d\tau\wedge\alpha)
 \end{equation}
 and
\begin{equation}\label{deriv2}
\kappa^2(\beta)=\psi \kappa(d\tau\wedge\underline{*}\beta)=
{\varphi}\psi(\underline{*}^2\beta)={\varphi}\psi(\beta)\,.
 \end{equation}
Consequently the operator $\kappa^2$ acts on an arbitrary form only by
multiplication by the  scalar factor ${\varphi}\psi$. Thus, we find the 
  reciprocity relation
 \begin{equation}\label{recip}
\kappa^2=(\varphi \psi)\,{\rm id}\,;
  \end{equation}
for a complete discussion of the physical meaning that can be given
to the reciprocity relation, see \cite{book}.

\subsection{Isotropic media}
We apply the definitions (\ref{const7}) and (\ref{const8}) to
 the electromagnetic fields appearing in (\ref{const5}), 
 \begin{equation}\label{const9}
   \kappa(d\tau\wedge E  )=\varphi \,\underline{*}E\,,\qquad
   \kappa(B)=\psi d\tau\wedge\underline{*}B\,.
    \end{equation}
Consequently (\ref{const5}) yields
\begin{equation}\label{const11}
   {\mathcal D}= {f_{\tt T}}\varphi\,\underline{*}E\,,\qquad
   {\mathcal H}=h_{\tt T}\psi\, \underline{*}B\,.
    \end{equation}
    These equations are reminiscent of the standard constitutive relations for 
    isotropic media. Thus
    we introduce the {\it electric permittivity}  and the
    {\it magnetic permeability} constants, respectively, 
    \begin{equation}\label{const12}
   \varepsilon= {f_{\tt T}}\varphi\,,\qquad \mu={h_{\tt T}}\,
   \frac 1\psi \,.
    \end{equation}
We can rewrite now the field equations (\ref{inhomMax}) and
(\ref{homMax}) as
   \begin{equation}\label{newMax}
    \left\{ \begin{array}{l}
 \psi\,\underline{d}(\underline{*}B)-f_{\tt T} \varphi\,\underline{*}\dot E =  j\,,\\ \\
  f_{\tt T}\varphi\,\underline{d}(\underline{*}E)=\rho \,,\end{array} \right.
    \qquad
    \left\{ \begin{array}{l}
 f_{\tt T}\,\underline{d}\,E-\dot{B}= 0\,,\\ \\
   \underline{d}\,B=0 \,.\end{array} \right.
    \end{equation}
    Observe that only one undefined  factor appears in these
    equations.
 \section{Energy-momentum current}
    \subsection{Premetric energy-momentum current}
    On a manifold endowed with a Minkowski metric, one is often dealing with a symmetric energy-momentum tensor.
    This tensor includes the products of the components of the fields $F$ and $H$ contracted by the metric tensor.
    Evidently such a construction is not suitable for the premetric approach which is dealing with  electromagnetic fields  on a bare differential manifold without a predefined metric structure. Moreover, the integration  cannot be applied directly to the tensor components due to  invariance argument. Since we need an integral over a region of a three-dimensional submanifold (the rest space), the integrand has to be a twisted  differential 3-form.

In the differential forms formalism, the  energy-momentum current  is
considered as a covector-valued 3-form $\Sigma_\alpha$.
In general, such a quantity has 16 independent components.
Already, in order to extract from $\Sigma_\alpha$ a symmetric energy-momentum tensor of 10 independent components, one needs a metric tensor. Indeed \cite{Itin:2001xz}, if a metric tensor $\eta^{\alpha\beta}$ is available,  one can define a 2-form $\eta^{\alpha\beta}e_\alpha\rfloor\Sigma_\beta$ of 6 independent components. In the case when this 2-form vanishes, the covector-valued 3-form $\Sigma_\alpha$ is equivalent to a symmetric tensor.
In the axiomatic electrodynamics formalism \cite{book},  the  energy-momentum current of
the electromagnetic field is postulated as
     \begin{equation}\label{em1aa}
       \Sigma_\alpha:= \frac 12\left[(e_\alpha\rfloor F)\wedge H-
       (e_\alpha\rfloor H)\wedge F\right]\,.
    \end{equation}
Here $e_\alpha$ is an arbitrary frame, not necessary a holonomic one. It is straightforward that for a coframe $\vartheta^\alpha$, which is dual to $e_\alpha$, the relation $\vartheta^\alpha\wedge\Sigma_\alpha=0$ holds.
This fact is equivalent to the tracelessness of the current $\Sigma_\alpha$. Consequently, the current (\ref{em1aa}) has 15 independent components in general.
\subsection{$(1+3)$ decomposition of the energy-momentum}
Relative to a chosen observer with a tangential vector ${\mathbf n}$, one can define an energy-momentum current
     \begin{equation}\label{em1aa-x}
       \Sigma:= \frac 12\left[({\mathbf n}\rfloor F)\wedge H-
       ({\mathbf n}\rfloor H)\wedge F\right]\,.
    \end{equation}
Evidently,  $ \Sigma=n^\alpha \Sigma_\alpha$.
    Since we are interested in the {\em energy} of the electromagnetic
    field, it is sufficient to discuss the current 
    $\Sigma$.

    Let us  decompose $\Sigma$ into time and space pieces.
 Because of ${\mathbf n}\rfloor d\tau=1$ and
    ${\mathbf n}\rfloor{\mathcal H}={\mathbf n}\rfloor{\mathcal D}={\mathbf n}\rfloor E={\mathbf n}\rfloor
    B=0$ (forms lie in the folio $\tau$),
    we find,  by using (\ref{Hdecomp}) and (\ref{Fdecomp}),
    \begin{eqnarray}\label{sigma0decomp}
      \Sigma&=&\frac 12 f_{\tt T}E\wedge (h_{\tt T}{d\tau\wedge\mathcal H}
      +{\mathcal D}) - \frac 12 h_{\tt T}{\mathcal H}\wedge
      (f_{\tt T}d\tau\wedge E + B ) \nonumber\\ &=&- h_{\tt
        T}f_{\tt T}d\tau\wedge E\wedge{\mathcal H} +\frac 12 f_{\tt
        T}\, {\mathcal D}\wedge E -\frac 12 h_{\tt T}\,{\mathcal H}\wedge
      B\,.
    \end{eqnarray}

 From the energy conservation law it follows that the 3-form $\Sigma_0$ must  have the $(1+3)$-decomposition 
of the type (\ref{decomp9})
 \begin{equation}\label{en-decomp}
       \Sigma= -d\tau\wedge \sigma+u\,,
    \end{equation}
 where the  3 form $u$ represents the electromagnetic energy,
 while the  2-form $\sigma$ is the electromagnetic energy flux. Both forms lie in the folio $\Sigma_\tau$. 
 Thus, the electromagnetic energy flux is 
  \begin{equation}\label{energy-flux}
\sigma= h_{\tt T}f_{\tt T} E\wedge{\mathcal H}\,
\end{equation}
and the energy is 
 \begin{equation}\label{em-energy}
 u=\frac 12 f_{\tt T}\, E\wedge {\mathcal D} -\frac 12 h_{\tt T}\,
 {\mathcal H}\wedge B\,.
 \end{equation}

There is a strong physical requirement: {\it The energy of the electromagnetic field
has to be positive}. The signs of the 3-forms  ${\mathcal D}\wedge E
$ and ${\mathcal H}\wedge B$ are  determined when the 
constitutive relations  (\ref{const11}) are used.  Indeed, 
\begin{equation}\label{energ1}
   u_{el}=\frac 12 f_{\tt T}\, E\wedge {\mathcal D}=
   \frac 12\varphi E\wedge \underline{*}E
  \end{equation}
  and
\begin{equation}\label{energ2}
u_{mag}=-\frac 12 h_{\tt T}\,{\mathcal H}\wedge B=
-\frac 12\,\psi\underline{*}B\wedge B\,.
    \end{equation}
  On a folio, the Euclidean metric is assumed. Thus the 3-forms
  $E\wedge \underline{*}E$ and
  $\underline{*}B\wedge B$
are positive. Consequently, both energy expressions (\ref{energ1})
and (\ref{energ2})
 are positive if and only if
\begin{equation}\label{energ-cond}
   \varphi>0 \,,\qquad \psi<0\,.
    \end{equation}

The energy conservation law takes the usual form, which is in  correspondence to the ordinary orientation of the boundary of a region, provided
\begin{equation}\label{flux2}
       d\Sigma_0=0 \qquad \Longleftrightarrow \qquad
       \underline{d}\,\sigma+\dot{u}=0\,.
    \end{equation}
From  the Maxwell system  (\ref{newMax}) and the constitutive relations
(\ref{const11}) we have in the source-free case 
\begin{equation}\label{newMax2}
\underline{d}\,\underline{*}\,B= f_{\tt T}\frac{\varphi}{\psi}\,\underline{*}\,\dot E\,,\qquad
 \underline{d}\,E=f_{\tt T}\dot{B}\,.
    \end{equation}
  Thus,
\begin{eqnarray}\label{check1}
\underline{d}\,\sigma&=&h_{\tt T}f_{\tt T}\underline{d}(E\wedge{\mathcal H})=
\psi f_{\tt T}\underline{d}(E\wedge\underline{*}\, B)=\psi f_{\tt T}(\underline{d}\,E\wedge\underline{*}\,B-E\wedge \underline{d}\,\underline{*}\,B)\nonumber\\
&=&\psi\dot{B}\wedge\underline{*}B-\varphi {\dot E}\wedge\underline{*} E
=- \frac 12\,\frac d{d\tau}\left(\varphi E\wedge\underline{*}E-\psi B\wedge\underline{*}B\right)=-\dot{u}\,.
  \end{eqnarray}
  Consequently, the conservation law is in  correspondence with the orientation of the boundary for arbitrary values of the sign factor $h_{\tt T}f_{\tt T}$. 
\subsection{Four types of electromagnetic media}
As a result of the consideration above, we remain with the undefined  sign factors $h_{\tt T}$ and
$f_{\tt T}$. Consequently, we have four possibilities for the signs, 
\begin{equation}\label{sign1}
h_{\tt T}=\pm 1\,,\qquad f_{\tt T}=\pm 1\,.
 \end{equation}
Recall that positivity  of the electromagnetic energy requires the condition (\ref{energ-cond}) for the parameters 
$\varphi$ and $\psi$. Using the definition of the 
of the   electric permittivity  and the
    magnetic permeability constants, respectively,
    \begin{equation}\label{const12-xxx}
   \varepsilon= {f_{\tt T}}\varphi\,,\qquad
   \mu={h_{\tt T}}\, \frac 1\psi \,,
    \end{equation}
    we can derive the following list for the possible signs for the
    electromagnetic constants:

\begin{center}
\begin{tabular}{|l|l|l|}
  \hline
$f_{\tt T}=+1\,,h_{\tt T}=-1$&$\varepsilon>0$&$\mu>0$\\
 \hline
$f_{\tt T}=-1\,,h_{\tt T}=+1$&$\varepsilon<0$&$\mu<0$\\
 \hline
$f_{\tt T}=+1\,,h_{\tt T}=+1$&$\varepsilon>0$&$\mu<0$\\
 \hline
$f_{\tt T}=-1\,,h_{\tt T}=-1$&$\varepsilon<0$&$\mu>0$\\
 \hline
  \end{tabular}
 \end{center}
Using the inequalities (\ref{energ-cond}), we are able to express the parameters $\varphi$ and $\psi$ and the sign factors via the   electric permittivity  and the magnetic permeability constants:
\begin{equation}\label{express1}
\varphi=|\varepsilon|\,,\qquad \psi=-\frac 1{|\mu|}
\end{equation}
and 
\begin{equation}\label{express2}
f_{\tt T}=\frac{\varepsilon}{|\varepsilon|}\,,\qquad h_{\tt T}=-\frac{\mu}{|\mu|}\,.
\end{equation}
    \section{Electromagnetic waves}
One of the principal facts of electromagnetic phenomena is
that the free electromagnetic field propagates by waves. Consider
the free Maxwell equations
 \begin{equation}\label{newMax1}
    \left\{ \begin{array}{l}
 \psi\,\underline{d}(\underline{*}B)-f_{\tt T} \varphi\,\underline{*}\dot E =  0\,,\\ \\
  f_{\tt T}\varphi\,\underline{d}(\underline{*}E)=0 \,,\end{array} \right.
    \qquad
    \left\{ \begin{array}{l}
 f_{\tt T}\,\underline{d}\,E-\dot{B}= 0\,,\\ \\
   \underline{d}\,B=0 \,.\end{array} \right.
    \end{equation}
Applying to the first equation the Hodge dual and the time
derivative and using the commutativity of these operations we obtain
\begin{equation}\label{wave1}
\psi\,\underline{*}\,\underline{d}\,\underline{*}\dot{B}- f_{\tt T} \varphi\,\ddot E = 0\,.
 \end{equation}
 From the first equation of the second system of (\ref{newMax1}), we have
\begin{equation}\label{wave2}
\dot{B}= f_{\tt T}\,\underline{d}\,E\,.
 \end{equation}
 Substituting it into (\ref{wave1}), we can rewrite it as
\begin{equation}\label{wave3}
\ddot E -\frac
{\psi}{\varphi}\underline{*}\,
\underline{d}\,\underline{*}\,\underline{d}E=0\,.
\end{equation}
Taking into account the equation $\underline{d}\,\underline{*}E=0$, we rewrite (\ref{wave5x}) as 
\begin{equation}\label{wave5x}
\ddot E +\frac {\psi}{\varphi}\triangle E=0\,.
\end{equation}
We introduced here the Laplace  operator
$\triangle=-dd^\dagger-d^\dagger d$ that in a three-dimensional Euclidean manifold acts on 1-forms as
\begin{equation}\label{wave6x}
\triangle
=-\underline{d}\,\underline{*}\,\underline{d}\,\underline{*}+
\underline{*}\,\underline{d}\,\underline{*}\,\underline{d}\,
\end{equation}
and on 2-forms as
\begin{equation}\label{wave6xx}
\triangle
=\underline{d}\,\underline{*}\,\underline{d}\,\underline{*}-
\underline{*}\,\underline{d}\,\underline{*}\,\underline{d}\,.
\end{equation}
Taking into account  (\ref{newMax1}), 
we are able to derive  that {\it all} four fields $E,B,{\mathcal D}$, and ${\mathcal
H}$ satisfy the same type of equation 
\begin{equation}\label{wave6-m}
\ddot {M} +\frac \psi{\varphi}\triangle M=0\,.
\end{equation}
 Since we assumed a
Euclidean metric in the folio $\Sigma_\tau$, we recognize that
in the case
\begin{equation}\label{wave8}
\frac\psi \varphi<0
\end{equation}
 the equation (\ref{wave6-m}) is hyperbolic and represents a wave equation. Thus, the corresponding four-dimensional optical metric is Minkowskian.

Observe that this result is independent on the  sign parameters
$f_{\tt T}$ and $h_{\tt T}$. Thus, it holds for any medium independently 
of the signs of the electromagnetic constants $\varepsilon$
and $\mu$.
Indeed, due to (\ref{express1}), we can rewrite (\ref{wave6-m}) as 
\begin{equation}\label{wave6}
 \ddot {M} -\frac 1{|\varepsilon\mu|}\triangle M=0\,.
\end{equation}
    \section{Lorentz force density}

    In this section, we turn to an additional and independent ingredient
    of  electromagnetic theory --- the Lorentz force.
    In the formalism of differential forms, the Lorentz force density
    is treated as a twisted covector-valued 4-form.
    Observe that this quantity has four independent components as in
    the ordinary vector description.
The Lorentz force is postulated \cite{book} as
    \begin{equation}\label{axiom2}
      {\mathcal F}_\alpha= (e_\alpha\rfloor F) \wedge J\,,
    \end{equation}
    where $e_\alpha$ denotes frame. 
    When the  $(1+3)$-decomposition of the field strength $F$
    is substituted, 
     the Lorentz force density (\ref{axiom2}) can be  decomposed according to
    \begin{equation}\label{lor5}
      {\mathcal F}_0=-f_{\tt T}\,E\wedge j\wedge d\tau\,
    \end{equation}
    and
    \begin{equation}\label{lor6}
      {\mathcal F}_\mu=\bigg[f_{\tt T}(e_\mu\rfloor E)\rho+
   (e_\mu\rfloor B)\wedge j\bigg]\wedge d\tau\,, \qquad
    \mu=1,2,3\,.
    \end{equation}
    Recall that we already have chosen the sign factor $ f_{\tt
    S}=+1$.

In particular, the Lorentz force  governs the law of attraction
and repulsion between charged particles. In order to
understand how the sign parameters are related to this law, we
consider a system of two static charged particles. In this case,
the Maxwell field equations (\ref{newMax}) show that the field
$E$ is static and satisfies the equations
\begin{equation}\label{lor7}
      {\underline d}E=0\,,\qquad
      {\underline d}\,{\underline *}E=\frac {f_{\tt T}}{\varphi}\,\rho_0\,.
    \end{equation}
    Here $\rho_0$ is the  charge density of the source. The sign of the solution of
    these equations, the sign of the field $E$,  is proportional
    to the sign of the factor $\frac {f_{\tt T}}{\varphi}$. When
    the solution is substituted into the electrostatic part of the
    Lorentz force,
\begin{equation}\label{lor8}
      {\mathcal F}_\mu=f_{\tt T}(e_\mu\rfloor E)\rho\wedge d\tau\,, \qquad
    \mu=1,2,3\,,
    \end{equation}
    the force density is generated in a special direction.
    We derive that the sign of the field ${\mathcal F}_\mu$ is
    proportional to the sign of the product $f_{\tt
    T}^2\varphi=\varphi$. Comparing to  ordinary dielectric
    materials (the first row in Table  1), we derive that in all media
    with a positive factor $\varphi$ the ordinary Dufay rule of attraction
     between opposite charges and repulsion between charges of
    the same sign holds true:
\begin{equation}\label{Dufay}
 \varphi>0\qquad \Longleftrightarrow \qquad {\textrm {Dufay rule}}.
    \end{equation}
    
    Consider now the pure magnetic contribution to the Lorentz
    force. Due to (\ref{newMax}), the magnetostatic case 
    can be described by the field $B$ satisfying  the equations
     \begin{equation}\label{mag-stat}
 \psi\underline{d}(\underline{*}B) =  j\,, \qquad
   \underline{d}\,B=0 \,.
    \end{equation}
    Consequently, the direction of the field $B$ is determined by
    the sign of the parameter $\psi$. A solution of
    (\ref{mag-stat}) is substituted now into the magnetic part of the
    Lorentz force,
  \begin{equation}\label{lor-mag}
      {\mathcal F}_\mu=
   (e_\mu\rfloor B)\wedge j\wedge d\tau\,, \qquad
    \mu=1,2,3\,.
    \end{equation}
Thus, also the direction of this force is determined by the sign
of the parameter $\psi$. Comparing to  ordinary media which are 
described by a negative parameter $\psi$, we derive that all cases
 with $\psi<0$ have the same behavior. In particular,
the Lenz rule (pulling of a ferromagnetic core into a solenoid
independently on the direction of the current) holds true:
\begin{equation}\label{lenz}
  \psi<0\qquad \Longleftrightarrow \qquad {\textrm {Lenz rule}}.
    \end{equation}
\section{Summary: Four types of media}
In this section we give a brief summary of the physical
properties of the four types of  media derived above. Recall that, in
the $4D$ formalism, all these media are described by the same
system of the field equations
\begin{equation}\label{sum1}
dF=0\,, \qquad dH=J\,.
\end{equation}
Moreover, we are dealing with the same isotropic constitutive
relation. 
Also the $4D$ expressions for
the energy-momentum  
\begin{equation}\label{sum2x}
  \Sigma_\alpha:= \frac 12\left[(e_\alpha\rfloor F)\wedge H-
       (e_\alpha\rfloor H)\wedge F\right]\,
\end{equation}
and for the Lorentz force density
 \begin{equation}\label{sum2xx}
      {\mathcal F}_\alpha= (e_\alpha\rfloor F) \wedge J\,,
    \end{equation}
 are the same. 
In all cases, the energy-momentum is conserved and its 
electric and magnetic energy parts are   positive. 
These parts can be written, using the absolute values of the electromagnetic parameters of the media, as 
\begin{equation}\label{sum3}
 u_{el}=\frac 12\,|\varepsilon| E\wedge \underline{*}E\,,\qquad 
u_{mag}=\frac 1{2|\mu|}B\wedge \underline{*}\,B\,.
\end{equation}
The standard Dufay and Lenz rules are satisfied. As we have shown, also here the absolute values of the electromagnetic parameters of the media appear.

 The difference between the four types of media originates from different expansions of the 
four-dimen\-sional fields relative to a chosen observer. Is it possible that all of these types present the same laws and 
transform one into another by redefinitions? In order to answer this question, let us consider the $4D$ invariants  of the electromagnetic field. They are defined as follows: 
\begin{equation}\label{sum4}
 I_1=F\wedge F=2f_{\tt T}d\tau\wedge E\wedge B\,,
\end{equation}
\begin{equation}\label{sum5}
 I_2=H\wedge H=
2h_{\tt T}\frac \varepsilon\mu\, d\tau\wedge {\underline *}E\wedge {\underline *}B\,,
\end{equation}
and
\begin{equation}\label{sum6}
 I_3=F\wedge H=
f_{\tt T}\varepsilon \,d\tau\wedge E\wedge{\underline *}E+h_{\tt T} \frac 1\mu\, d\tau\wedge B\wedge{\underline *}B
\end{equation}
  Since the expressions $f_{\tt T}\varepsilon$ and $h_{\tt T} \mu$ are the same in all four models, the invariant $I_3$ is also the same. However, the invariants  $I_1$ and $I_2$  are different. Thus, they can  be used  for  the separation of the physical features of our models. 
  
\begin{table}[h]
\begin{center}
\begin{tabular}{|l|l|l|l|l|}
  \hline
  &&&&\\
&$f_{\tt T}=+1$&$f_{\tt T}=-1$&$f_{\tt T}=+1$&$f_{\tt T}=-1$\\
&$h_{\tt T}=-1$&$h_{\tt T}=+1$&$h_{\tt T}=+1$&$h_{\tt T}=-1$\\
&&&&\\ \hline
 &&&&\\
  parameters&$\varepsilon>0\,,\quad\mu>0$&$\varepsilon<0\,,\quad\mu<0$&$\varepsilon>0\,,\quad\mu<0$&$\varepsilon<0\,,\quad\mu>0$\\
   &&&&\\
  \hline &&&&\\
 $H=$ &$-d\tau\wedge {\mathcal H}+\,{\mathcal D}$&$d\tau\wedge {\mathcal H}+\,{\mathcal D}$&$d\tau\wedge {\mathcal H}+\,{\mathcal D}$&$-d\tau\wedge {\mathcal H}+\,{\mathcal D}$\\
&&&&\\
  \hline &&&&\\
  $F=$ &$d\tau\wedge E+B$&$-d\tau\wedge E+B$&$d\tau\wedge E+B$&$-d\tau\wedge E+B$\\
&&&&\\
  \hline &&&&\\
Maxwell-1&$-\underline{d}\,{\mathcal H}-\dot{\mathcal D}=j$&$\underline{d}\,{\mathcal H}-\dot{\mathcal D}=j$&$\underline{d}\,{\mathcal H}-\dot{\mathcal D}=j$&$-\underline{d}\,{\mathcal H}-\dot{\mathcal D}=j$\\
 &&&&\\ \hline &&&&\\
 Maxwell-2&$\underline{d}\,\mathcal D=\rho$&$\underline{d}\,\mathcal D=\rho$&$\underline{d}\,\mathcal D=\rho$&$\underline{d}\,\mathcal D=\rho$\\
 &&&&\\ \hline &&&&\\
 Maxwell-3& $\underline{d}\,E-\dot{B}=0$& $\underline{d}\,E+\dot{B}=0$& $\underline{d}\,E-\dot{B}=0$&
 $\underline{d}\,E+\dot{B}=0$\\
&&&&\\   \hline &&&&\\
Maxwell-4& $\underline{d}\,B=0$&  $\underline{d}\,B=0$&  $\underline{d}\,B=0$&  $\underline{d}\,B=0$\\
 &&&&\\ \hline
 &&&&\\
invariants& $I_1\,,\qquad I_2$&  $-I_1\,,\qquad -I_2$&  $I_1\,,\qquad -I_2$&  $-I_1\,,\qquad I_2$\\
 &&&&\\ \hline
  \end{tabular}
   \end{center}\end{table}
   In Table 1 we present the main features of the four types of isotropic electromagnetic media.
Notice that the sign in front of $\underline{d}\,{\mathcal H}$ in the first Maxwell equation is opposite to the sign ordinarily used. This is due to our form of the $1+3$ decomposition of $H$. 

\section{Conclusions and discussion}
Minkowski's idea on the four-dimensional space as a proper physical reality includes a $4D$ reformulation of Maxwell's electrodynamics. We start with a premetric $4D$ construction which is explicitly covariant and  does not  even involve a metric tensor. Alternatively, the metric with Minkowskian signature may be incorporated into the structure of the field equations and the constitutive relations. This structure is unique, provided the following  physically meaningful conditions are assumed:
  \begin{itemize}
   \item[i] Charge conservation;
   \item[ii] flux conservation; 
\item[ii] positivity  of the electric and magnetic  energy;
\item[iii] energy-momentum conservation;
\item[iv] local linear constitutive law which we restrict to the isotropic form.
 \end{itemize}
Does this mean that the backward $(1+3)$-splitting of the general covariant Maxwell system is unique? 
We show in this paper that the answer is negative. In the framework of the conditions listed above, we derive that four different sets of the three-dimensional Maxwell equations correspond to the same four-dimensional system. These sets are different because of the signs of the values of the electromagnetic constants. So we can treat them as four different electromagnetic media. The main features of  standard electrodynamics remain true for all these media. In particular, the electric and magnetic energies are positive and the standard direction of the Lorentz force (Dufay's and Lenz's rules) is preserved. However, these models  differ from one  another in particular by the $4D$ electromagnetic invariants. 

Recently  media with non-positive electromagnetic parameters are studied intensively
\cite{Veselago}--\cite{Sihvola}. Moreover,  such materials (called  {\it metamaterials}, or {\it left-handed materials}) are already manufactured and even proposed to be useful in technology. 
 However, the description of such materials is sophisticated. In particular, these metamaterials are treated sometimes as electromagnetic media in which the energy of the electromagnetic field is negative. Note that the theoretical approach used in  metamaterials is completely different from ours. In particular, the Maxwell equations which were developed and tested only for $\varepsilon>0,\mu>0$ are used in the situations where such assumptions do not hold. It is not surprising that  contradictions to the basic laws of physics emerge. 

Alternatively, in our approach the same type of non-ordinary media are derived from the unique $4D$ system and the basic physical laws are preserved.   
\section*{Acknowledgment}
 Our  acknowledgment goes to Friedrich Hehl for inviting this paper and for his most helpful comments.  


\section*{References}
  \begin{thebibliography}{99}
      \bibitem{Minkowski} H.~Minkowski, {\it Space and time} in {\it The
  Principle of Relativity} (Dover Pub., New York, 1923).
  \bibitem{Post62} E.J.~Post, {\it Formal Structure of Electromagnetics
    -- General Covariance and Electromagnetics} (North Holland, 
  Amsterdam, 1962, and Dover, Mineola, New York, 1997).
  \bibitem{ROH2002} G.F.~Rubilar, Yu.N.~Obukhov, and F.W.~Hehl, {\em
    General covariant Fresnel equation and the emergence of the light
    cone structure in pre-metric electrodynamics}, Int.\ J.\ Mod.\ 
  Phys.\ {\bf D11},  1227--1242 (2002) [arXiv:gr-qc/0109012v2].
  
\bibitem{gentle} F.W.\ Hehl and Yu.N.\ Obukhov, {\em A gentle
    introduction into the foundations of classical electro\-dynamics:
    Meaning of the excitations $({\mathcal D},{\mathcal H})$ and the field
    strengths $(E,B)$} [arXiv: physics/0005084]  (2000).
      
\bibitem{Obukhov:2000nw}
  Y.~N.~Obukhov, T.~Fukui and G.~F.~Rubilar,
  {\em Wave propagation in linear electrodynamics,}
  Phys.\ Rev.\  D {\bf 62}, 044050 (2000).
\bibitem{Lammerzahl:2004ww}
  C.~L$\ddot{\rm a}$mmerzahl and F.~W.~Hehl,
 {\em Riemannian light cone from vanishing birefringence in premetric vacuum
  electrodynamics,}
  Phys.\ Rev.\  D {\bf 70}, 105022 (2004).


\bibitem {HIO} F.W. Hehl, Y. Itin, Yu.N. Obukhov, {\em Recent developments in premetric classical electrodynamics}, Proceedings of the 3rd Summer School in Modern Mathematical Physics, 20-31 August 2004, Zlatibor, Serbia and Montenegro B. Dragovich et al., eds., SFIN (Notebooks on Physical Sciences) XVIII: Conferences, A1 (2005) 375-408 (Institute of Physics: Belgrade, 2005) [arXiv.org/physics/0610221].

 \bibitem{book} F.W.~Hehl and Yu.N.~Obukhov, {\it Foundations of
        Classical Electrodynamics: Charge, Flux, and Metric}
      (Birkh\"auser, Boston, MA, 2003).
      
    \bibitem{Itin:2004qr}
      Y.~Itin and F.~W.~Hehl,
      {\em Is the Lorentz signature of the metric of spacetime electromagnetic in
      origin?,}
      Annals Phys. (N.Y.)\  {\bf 312}, 60 (2004).
      
\bibitem{Bini:1995hx}
  D.~Bini, P.~Carini and R.~T.~Jantzen,
 {\em Relative observer kinematics in general relativity,}
  Class.\ Quant.\ Grav.\  {\bf 12}, 2549 (1995).

\bibitem{Hehl:2004jn}
  F.~W.~Hehl and Y.~N.~Obukhov,
  Gen.\ Rel.\ Grav.\  {\bf 37}, 733 (2005)
  [arXiv:physics/0407022].
\bibitem{Itin:2001xz}
  Y.~Itin,
  {\em Coframe energy-momentum current. Algebraic properties,}
  Gen.\ Rel.\ Grav.\  {\bf 34}, 1819 (2002)
  [arXiv:gr-qc/0111087].

\bibitem{Veselago} V.G. Veselago  {\em The electrodynamics of substances with simultaneously negative values of $\varepsilon$  and $\mu$} Sov. Phys. Usp. {\bf {10}}, 509 (1968).

\bibitem {shelby} R. Shelby, D. R. Smith and S. Schultz, {\em Experimental verification of a negative index of refraction}, Science {\bf {292}}, 77 (2001).

\bibitem{Lakhtakia} R.A. Depine and A. Lakhtakia, {\em A new condition to identify isotropic dielectric-magnetic materials displaying negative phase velocity}, Microwave and Optical Technology Letters {\bf 41}, 315 (2004).

\bibitem{Sihvola} A.H. Sihvola, {\em Metamaterials in electromagnetics}, Metamaterials {\bf 1}, 2 (2007).
    \end{thebibliography}
    \end{document}